\documentclass[a4paper,11pt]{article}
\pdfoutput=1 

\usepackage{jcappub} 

\usepackage[T1]{fontenc} 

\usepackage{amsfonts}
\usepackage{amsmath}
\usepackage{amssymb}
\usepackage{graphicx,color}
\usepackage{float}
\usepackage{hyperref}
\usepackage{subfigure}
\usepackage{dcolumn}
\usepackage{soul}
\usepackage{verbatim}
\usepackage{bm}  

\title{The early universe is \textit{ACT}-ing \textit{warm}} 

\author[a]{Arjun Berera,}
\author[a,b,1]{Suddhasattwa Brahma\note{Corresponding author.},}
\author[a]{Zizang Qiu,}
\author[c]{Rudnei O. Ramos}
\author[c]{and Gabriel S. Rodrigues} 

\affiliation[a]{Higgs Centre for Theoretical Physics, School of Physics and Astronomy, University of Edinburgh, Edinburgh EH9 3FD, UK}
\affiliation[b]{Physics and Applied Mathematics Unit, Indian Statistical Institute, 203 B.T. Road, Kolkata 700108, India}

\affiliation[c]{Departamento de Fisica Teorica, Universidade do Estado do
  Rio de Janeiro, 20550-013 Rio de Janeiro, RJ, Brazil }

\emailAdd{ab@ed.ac.uk}
\emailAdd{suddhasattwa.brahma@gmail.com}
\emailAdd{Zizang.Qiu@ed.ac.uk}
\emailAdd{rudnei@uerj.br}
\emailAdd{gabriel.desenhista.gr.gr@gmail.com}

\abstract{The recently released data from the \textit{Atacama Cosmology Telescope} (ACT) confirms that the primordial scalar spectrum is extremely flat. This, together with current upper bounds on the tensor-to-scalar ratio, implies that the simplest models of inflation coming from particle physics (for instance, a minimally-coupled scalar with monomial potentials) need additional ingredients in order to make them compatible with observations. Instead of invoking arbitrary new couplings or new interactions that are not protected symmetries, we argue that dissipation of the inflaton field with the radiation bath should be added as a new physical principle. Accordingly, we show that warm inflation provides the correct paradigm to explain the current observations, given very natural choices of dissipative terms. The model analyzed here has mirror and $Z_4$ symmetries, which explicitly protect the inflaton potential from large quantum and thermal corrections. We use a recent precision numerical code designed for warm inflationary perturbations, improving on the determination of the cosmological observables previously obtained for such models.}
\begin{document}
\maketitle

\section{Spoiled for choice, yet not a single `principled' one}

Our current understanding of the early universe paints a picture that is extraordinarily simple. The background spacetime is believed to have undergone an exponential expansion, commonly termed as \textit{inflation}. This solves the puzzles of standard Big Bang cosmology such as why the sky looks the same in all directions or why the spatial curvature of the universe seems to be zero. However, it not only explains why the universe seems homogeneous and isotropic on large scales, but also provides the seeds for the small temperature anisotropies that we observe in the Cosmic Microwave Background (CMB) radiation (or as late-time inhomogeneities in the Baryonic Acoustic Oscillations or Large Scale Structure data). 

The main prototype for inflation is a scalar field slowly rolling down an almost flat potential, such that the potential energy of the field dominates over its kinetic energy, and sources the Friedmann equation. The Hubble radius remains approximately constant during this phase, leading to an increase of the size of the universe by a factor of $e^{3N}$ if inflation lasted for $N$ $e$-folds. This results in a supercooled universe which is practically empty, and one needs a phase of reheating/preheating where the energy density of the inflaton is released by creating a shower of particles in radiation form. 

The above story is simple enough that theorists have successfully built a zoo of potentials ($\sim\mathcal{O}(100)$ \cite{Martin:2013tda}) for single field models of inflation. However, if we want to go beyond pure phenomenology to construct a particle-physics motivated models of inflation, one finds that the main restriction is that the potential must remain sufficiently flat for long enough for inflation to take place. In other words, not only should the classical potential have a very flat shape, it needs to be protected from quantum corrections. This is commonly referred to as the eta-problem in inflation: How do we keep quantum radiative corrections from generating a large mass term for the inflaton? Note that the theoretical obstruction comes from noting that the potential should be sensitive to Planck-suppressed operators, the absence of which would be an extreme fine-tuning problem. However, more importantly, current observations tell us that the primordial density perturbations have an adiabatic, Gaussian and near scale-invariant spectrum. This requires that the effective field theory (EFT) of inflation is that of a light scalar field, which is precisely what the above-mentioned eta problem makes it difficult to realize. 

One way to achieve the above would be protecting the flatness of the potential via symmetries, such as supersymmetry (which would cancel the bosonic and fermionic contributions to the mass) or having an approximate shift symmetry (see, for instance, \cite{Baumann:2014nda}). The first suggestion works for highly fine-tuned superpotentials \cite{McAllister:2007bg}, which just kicks the naturalness problem to a different starting point. Axion fields do have weak shift symmetries \cite{Freese:1990rb}, but these are ruled out from both theoretical (super-Planckian decay constants are ruled out by the Weak Gravity Conjecture \cite{Rudelius:2015xta, Heidenreich:2015wga}) as well as from observational constraints \cite{Stein:2021uge}. For inflationary models which require a large field excursion, the sensitivity of the potential to Planckian physics is enhanced, further worsening this problem.  The broader point is that since there are no global symmetries in Quantum Gravity \cite{Banks:2010zn}, this means all such models employing some symmetry would require a degree of fine-tuning to make them work. A recent, systematic manner of characterizing these problems is the so-called `Swampland program' \cite{Palti:2019pca, Bedroya:2019tba}.

What would be desirable would be to have a principled model such as (modified or extended) Higgs inflation \cite{Bezrukov:2007ep, Giudice:2010ka, Bauer:2010jg, Barbon:2015fla, Ema:2017rqn, Shaposhnikov:2020gts, Bojowald:2020eem}. The non-minimally coupled Higgs is a variation  \cite{Rubio:2018ogq, Calmet:2016fsr} of the Starobinsky model \cite{Starobinsky:1980te}, which is a prototype of the so-called $\alpha$-attractors \cite{Kallosh:2013yoa}. The latest ACT results~\cite{ACT:2025tim,ACT:2025fju}, which in combination with the Planck \cite{Planck:2018jri,Planck:2018vyg} and DESI data \cite{DESI:2024mwx, DESI:2024uvr}, present the state-of-the-art for the CMB primordial spectrum. It shows that Starobinski inflation is now disfavored at 2$\sigma$ (at least in its simplest implementation~\cite{ACT:2025tim}) since the density spectrum is even flatter than we had anticipated earlier. This, coupled with the fact that we have not observed primordial gravitational waves yet, implies that most of the particle-physics inspired models of single-field cold inflation have been ruled out by CMB data. In particular, simple renormalizable models such as $\phi^4$ or $\phi^2$ potentials, characterizing chaotic inflation, or even the non-minimally coupled Higgs models from this paradigm have now all either been ruled out or are strongly disfavoured by data.

This work presents several advancements over previous studies in multiple respects:  
\begin{enumerate}
\item We begin by performing a new analysis of the warm little inflaton (WLI) model originally introduced in Ref.~\cite{Bastero-Gil:2016qru}, employing the recently released \texttt{WI2easy} code~\cite{Rodrigues:2025neh}. This code enables a more efficient and precise computation of both the background dynamics and curvature perturbations, improving upon the accuracy of earlier numerical treatments. For the benchmark parameter set considered in Ref.~\cite{Bastero-Gil:2016qru}, corresponding to a dissipation ratio $ Q = \Upsilon / (3H) = 0.27 $, we obtain updated values for the spectral index and tensor-to-scalar ratio, $ n_s = 0.960 $ and $ r = 9.5 \times 10^{-4} $, respectively. These should be compared with the previous estimates $ n_s = 0.964 $ and $ r = 8.0 \times 10^{-4} $ reported in Ref.~\cite{Bastero-Gil:2016qru}. The variation in $ n_s $ is already at the level of sensitivity of the precision data from ACT and \textit{Planck}.
\item In addition, we re-examine, using the \texttt{WI2easy} framework, the results of Ref.~\cite{Bastero-Gil:2019gao}, where dissipation arises from direct interactions between the inflaton and scalar fields, in contrast with the fermionic interactions originally considered in Ref.~\cite{Bastero-Gil:2016qru}.
\item Beyond improving the precision in determining the power spectrum for warm inflation, we evaluate all quantities at the Hubble radius crossing point $ N_* $ for the pivot scale. To this end, we follow the post-inflationary dynamics up to the onset of effective radiation domination, where the equation of state approaches $ w = 1/3 $. This procedure allows for a more accurate estimation of observable quantities. In particular, while Refs.~\cite{Bastero-Gil:2016qru} and \cite{Bastero-Gil:2019gao} presented results at a fixed number of e-folds, $ N_* = 60 $, the more precise analysis performed here yields $ N_* = 58.27 $ for the benchmark case studied in Ref.~\cite{Bastero-Gil:2016qru} ($ Q_* = 0.27 $) and $ N_* = 55.86 $ for the benchmark case studied in Ref.~\cite{Bastero-Gil:2019gao} ($ Q_* = 100 $).
\item {}Furthermore, we extend the analysis to a unified WLI scenario that simultaneously incorporates both fermionic and scalar dissipative channels, performing a detailed study of the resulting dynamics and predictions. This is the first time that this extended and more complete model is studied.
\item All numerical analyses for the three models have been contrasted with the most recent ACT CMB data, as well as with the earlier \textit{Planck}/BICEP measurements.
\item We also provide updated results for the running of the spectral index, which recent ACT observations suggest may take positive values. We examine this running separately for the fermionic, scalar, and combined interaction cases within the full WLI framework. Our results indicate that the recent CMB data from ACT show a clear preference for the strong dissipation regime of warm inflation.
\item {}Finally, we present, for the first time, a dedicated study of the possible thermalization of the inflaton in the extended WLI model, identifying parameter regions that favor such thermalization and discussing the resulting implications for the power spectrum and other observable quantities.
\end{enumerate}

Throughout this paper, we employ the natural units: the
speed of light,  Planck's constant, and Boltzmann's constant are all
set to $1$, $c=\hbar=k_B=1$, respectively.  We consider a spatially flat homogeneous and isotropic
background metric with scale factor $a(t)$, where $t$ is physical
time. Dimensional quantities are expressed in terms of the reduced Planck mass,
$M_{\rm Pl} = (8\pi G)^{-1/2} \simeq 2.44\times 10^{18}$ GeV, and where $G$ is Newton's gravitational constant.
The Hubble expansion rate is
$H(t) = {\dot{a}(t)}/a(t)$ and overdots mean derivative
with respect to time.

The remainder of this paper is organized as follows. In section~\ref{wireview}, we briefly review the warm inflation (WI) dynamics and the role of dissipation. In section~\ref{model}, we explain the microphysics of the WI model used in our analysis. In section~\ref{results}, we give all our numerical results and their connection to the recent ACT data.
Our conclusions are given in section~\ref{conclusions}. {}Finally, in the appendix~\ref{appA} we analyze the possibility of thermalization of the inflaton excitations in the model we have studied.

\section{A new physical input: Dissipative corrections}
\label{wireview}

The above discussion implicitly ignores the interactions of the inflaton with other matter components during the quasi-dS expansion of the universe or, at best, assumes that these interactions only affect the potential through quantum corrections. However, one should also include dissipative effects that typically stem from interactions of the inflaton field. If the radiation production rate is non-negligible during inflation, this would compensate the super-cooling due to the exponential expansion and would not result in an empty universe.  This is what has been shown to be the case in standard models of the WI paradigm \cite{Berera:1995ie}, which results in a radiation bath that is in quasi-thermal equilibrium. As there exists a radiation bath, the energy of the inflaton field can be dumped into this environment continuously through dissipative terms provided the microphysics causing dissipation has a characteristic time-scale that is much larger than the local expansion rate of the background spacetime \cite{Berera:1996nv,Berera:1998gx}. As is obvious from this description, there is no need for reheating in WI as it allows a smooth and gradual phase transition from exponential expansion to a radiation-dominated standard big-bang cosmology \cite{Berera:1996fm}.

Unlike in cold inflation, where the height of the potential determines the energy density of the universe while its slope fixes the slow-roll dynamics, there is an additional friction term in WI due to the radiation bath that keeps the universe warm despite the supercooling quasi-dS expansion. The background dynamics in WI is still dominated by the potential energy of the inflaton field, $V$. However, the inflaton field dynamics is not isolated and there exists the radiation energy density $\rho_r^{1/4} > H$, where $H$ is the approximately constant Hubble parameter during inflation. The resulting background dynamics in WI, for the inflaton and radiation energy density, modified by the presence of a dissipation term
$\Upsilon$, are, respectively \cite{Berera:1995ie}
\begin{eqnarray}
&&\ddot \phi + (3 H + \Upsilon) \dot \phi + V_{,\phi}=0,
\label{eqphi}
\\
&&\dot \rho_r + 4 H \rho_r = \Upsilon \dot \phi^2,
\label{eqrhor}
\end{eqnarray}
where $\Upsilon$ is the dissipation coefficient 
(for reviews on its derivation from first principles
quantum field theory, see, \textit{e.g.} \cite{Berera:2008ar,Kamali:2023lzq} and references there in) and it can be, in general, a function of both the temperature $T$ of the generated thermal radiation bath and on the inflaton amplitude $\phi$.

{}For WI to take place, the energy density of the inflaton field must dominate over that of the radiation field. On introducing the dissipation coefficient $Q= \Upsilon/(3H)$, one can rewrite standard slow-roll parameters as 
\begin{eqnarray}
    \epsilon_V &=& \frac{M_{\rm Pl}^2}{2} \left(\frac{V'}{V}\right)^2 < 1+ Q\,,\\
    \eta_V &=& M_{\rm Pl}^2 \left(\frac{V''}{V}\right) < 1+Q,
\end{eqnarray}
where $M_{\rm Pl}$ is the reduced Planck mass and $Q$ is the dissipation ratio $Q= \Upsilon/(3H)$. 
It is clear that if WI allows for strong dissipation, $Q\gg 1$, then the problem of requiring that the inflaton mass remains negligible during inflation is automatically solved. Put another way, the strong dissipation allows for a slow-roll trajectory by proving additional friction even if quantum corrections somewhat spoil the flatness of the potential. 

The presence of the dissipation term in WI not only modifies the background dynamics, but also strongly affects the perturbations, with in particular the inflaton and radiation perturbations becoming coupled due to the temperature dependence on $\Upsilon$.  The
scalar primordial perturbations for warm inflation
in the thermally induced case were obtained here
\cite{Berera:1995wh,Berera:1999ws}. Thus, the
scalar perturbation power spectrum in WI is modified with respect to the CI case and it
 can be written as~\cite{Ramos:2013nsa,Benetti:2016jhf}
\begin{eqnarray}
P_{\mathcal R}(k/k_*)=\left(\frac{H^2}{2\pi\dot\phi}\right)^2\mathcal{F}(k/k_*),
\label{full-power}
\end{eqnarray}
where all quantities are meant to be evaluated at 
Hubble radius crossing scale $k=aH$.  In eq.~(\ref{full-power}), 
$\mathcal{F}(k/k_*)$ is defined as
\begin{eqnarray}
\mathcal{F}(k/k_*)\equiv\left(1+2n_*+\frac{2\sqrt3\pi
  Q}{\sqrt{3+4\pi Q}}\frac{T}{H}\right)G(Q). \nonumber \\
\label{Fk}
\end{eqnarray}
Here the last term in the parenthesis arises from the thermalized heat bath inducing fluctuations of the inflation field through the fluctuation-dissipation dynamics. In addition, the inflaton field can also become thermalised and $n_*$ accounts for the possibility of a thermal distribution of the inflaton field with a
Bose-Einstein distribution $n_* \equiv n_{\rm BE} = 1/[\exp(H/T)-1]$ when fully thermalized (for intermediate cases,
see, e.g. ref.~\cite{Bartrum:2013fia,Bastero-Gil:2017yzb}).
In the appendix~\ref{appA}, we discuss the possibility of thermalization 
of the inflaton and give a justification of when it is appropriate to consider $n_* \equiv n_{\rm BE}$ in eq.~(\ref{Fk}). Finally the first term in the above
parenthesis, the $1$, reflects the  contribution of the ground state vacuum fluctuations.
In eq.~(\ref{Fk}), the function $G(Q)$ models the
effect of the coupling of the inflaton and radiation fluctuations and which can only be determined numerically by solving the coupled system of perturbation equations in WI~\cite{Graham:2009bf, Bastero-Gil:2011rva,Bastero-Gil:2014jsa} (see also 
refs.~\cite{Montefalcone:2023pvh,Rodrigues:2025neh}
for public codes that are dedicated to evolve the full set of perturbation equations in WI and to numerically determine $G(Q)$ for different forms of dissipation coefficients). 
Equation~(\ref{full-power}) can be expanded in terms of the small-scale $k$-dependence as
\begin{eqnarray}
P_{\mathcal R}(k/k_*) \simeq  A_s
\left(\frac{k}{k_*}\right)^{n_s-1 + \frac{\alpha_s}{2} \ln(k/k_*)
  +\cdots},
\label{PR}
\end{eqnarray}
where $A_s$ is the scalar amplitude, $n_s$ is the scalar tilt,
\begin{equation}
n_s -1 =   \frac{d \ln P_{{\cal R}}(k/k_*) }{d \ln(k/k_*)}\Bigr|_{k\to k_*} ,
\label{eq:n}
\end{equation}
and
$\alpha_s$ is the running (the ellipses in eq.~(\ref{PR}) denote higher order running terms, which we neglect here),
\begin{equation}
\alpha_s=   \frac{dn_s(k/k_*) }{d \ln(k/k_*) }\Bigr|_{k\to k_*}.
\label{dns}
\end{equation}
In WI, expressions for the tilt and running can be obtained explicitly by using eq.~(\ref{full-power}) (see also, \textit{e.g.} ~\cite{Das:2022ubr}).

The Planck+ACT+DESI data (P-ACT-LB hereafter) gives the best fit value of $n_s = 0.974 \pm 0.003$, which clearly shows that the spectrum is even more scale invariant. If one adds in the BICEP/\textit{Keck} (P-ACT-LB-BK18 hereafter) data~\cite{BICEP:2021xfz, BICEPKeck:2024stm}, one can turn this into a constraint on the second slow-roll parameter $\eta_V \sim \mathcal{O}(10^{-3})$~\cite{ACT:2025fju, ACT:2025tim}, which is very close to zero. Thus, the almost scale-invariance of the spectrum can be seen to make the eta-problem of cold inflation even worse, requiring extreme fine-tuning of the alpha-attractor models to include further arbitrary parameters of non-minimal couplings of the inflaton to gravity \cite{Kallosh:2025rni, Kallosh:2013tua}. On the other hand, this is not required in WI. For the strong dissipation regime, the value of $n_s$ is naturally led to larger values closer to one. The physical reason behind this is that a given wavelength mode attains its constant value slightly before it exits the horizon, whereas the dissipative terms slow down the evolution of the field.  In addition to slightly larger values for the scalar tilt, the recent ACT data also indicates a greater preference for a  positive running, with $\alpha_s = 0.0062 \pm 0.0052$, while the Planck results pointed toward a sllightly
negative value~\cite{Planck:2018vyg,Planck:2018jri}, $\alpha_s=-0.0041 \pm 0.0067$. As we shall see, warm inflation can easily accommodate both of these results, in particular in the so-called strong dissipative regime $Q>1$.

There is both a conceptual and practical importance of having warm inflation in the strong dissipative regime.
Although quantum fluctuations are omnipresent in all models of inflation, eq.~(\ref{full-power}) shows that the thermal and dissipation effects can dominate over them in WI and are primarily responsible for the sourcing adiabatic curvature perturbations. More interestingly, in the strong dissipative regime, where $Q > 1$, these fluctuations freeze out even before they become superhorizon leading to an enhancement in the amplitude of the power spectrum. Since there is no such modification to the spectrum of the primordial tensor modes, this means that WI models would always lead to a suppression of the tensor-to-scalar ratio $r$ (noticing that dissipative effects to the gravitons are tiny for any temperature below the Planck scale). This implicitly is another reason why the data appear to favor WI models, as the upper bound on $r$ is now constrained to be below $0.036$ \cite{BICEP:2021xfz}. This will be further constrained in the near future by various missions such as LiteBird, BICEP/\textit{Keck}, CMB-S4 etc. How large the value of $r$ we find would be the smoking gun in deciding between weak and strong dissipative regimes of WI, and would be crucial as an indication of whether the `natural' models of WI, parametrized by a large value of $Q$, are preferred or not.

\section{It is time to \textit{ACT} now}
\label{model}

P-ACT-BL has solidified the trend of pushing $n_s$ to higher values while combining the BK-18 results on a tiny upper bound on the tensor-to-scalar ratio. We would have to wait until the next data release of South Pole Telescope (SPT) to see if this trend of increasing $n_s$ while other missions will further constrain the upper bound on the amplitude of primordial gravitational waves. Our main message is that one does need to keep on fine-tuning existing cold inflation models by further adding parameters to account for future observations. Instead, a principled approach would be to see what would happen if we were to add the dissipative corrections coming from WI and see what they predict the typical values for $n_s$ and $r$ are. 

The warm inflation model building program has focused on models built with the types of fields and symmetries that are as close as possible to those already found
in the Standard Model, and have thus been experimentally observed. Developing such
a goal has been possible due to two key steps.  The first, we made in
the early years of the development of WI was the recognition
that when the dissipative coefficient $\Upsilon$ is larger than
the Hubble scale $H$, \textit{i.e.} $\Upsilon > 3H$, it is possible in
this strong dissipative regime for the inflaton mass
to be larger than the Hubble scale, $m_{\phi} > H$, and
the inflaton field amplitude to remain below the Planck scale
$\phi < M_{\rm Pl}$ \cite{Berera:1999ws,Berera:2003yyp}.
Such a regime is then protected from quantum gravity corrections arising in higher-dimensional operators and is free of uncontrollable infrared problems.  This is the minimal regime necessary to be consistent with the quantum field theory understood by the Standard Model.  The second key step was recognizing, after much research by some of
us~\cite{Bartrum:2013fia,Bastero-Gil:2009sdq}, that for WI it is not only Supersymmetry that best protects the inflaton mass from radiative and thermal corrections, but rather there can be other symmetries,
particularly those reminiscent of beyond the standard model Higgs phenomenology, leading to the first WI model based on this symmetry, the Warm Little inflaton (WLI) model~\cite{Bastero-Gil:2016qru}.  These two steps created a new path for inflation model building and the one followed in most WI model construction today (for a more detailed discussion, see~\cite{Berera:2023liv,Kumar:2024hju}).

\subsection{The warm little inflaton model}

In all of our analysis, we will consider a generalization of the WLI model, which unifies the models first introduced in refs.~\cite{Bastero-Gil:2016qru,Bastero-Gil:2019gao}
(see also ~\cite{Bastero-Gil:2018uep,Ferraz:2023qia} for applications and other generalizations).
The WLI model is characterized by an interaction Lagrangian density given by
\begin{eqnarray} 
\mathcal{L}_{\rm int} &=& -{g_{\phi\psi}\over
  \sqrt{2}}(\phi_1+\phi_2)\bar\psi_{1L} \psi_{1R} + i{g_{\phi\psi}\over
  \sqrt{2}}(\phi_1-\phi_2)\bar\psi_{2L} \psi_{2R}
\nonumber \\
&-&{1\over2}g_{\phi\chi}^2|\phi_1+\phi_2|^2|\chi_1|^2-
          {1\over2}g_{\phi\chi}^2|\phi_1-\phi_2|^2|\chi_2|^2\nonumber \\
&-& h_{\sigma\psi}\sigma \sum_{i=1,2}\left(
\bar{\psi}_{iL}\psi_{\sigma R}+ \bar{\psi}_{\sigma
  L}\psi_{iR}\right) \nonumber \\
&+&\sum_{\substack{i\neq
    j=1,2}}\left(h_{\chi\psi}\chi_i^\dagger\bar\psi_{\sigma L}\psi_{\sigma R}+\mathrm{h.c.}\right),
\label{WLI}
\end{eqnarray} 
where $\phi_1$ and
$\phi_2$ are two complex scalar fields, with a symmetry breaking potential,
\begin{equation}
V(\phi_1,\phi_2) =  \sum_{i=1,2}   \frac{\lambda_i}{4} \left(|\phi_i|^2-\frac{M^2}{2}\right)^2,
\label{pot}
\end{equation}
and vacuum expectation values,
$\langle \phi_1\rangle= \langle
\phi_2\rangle \equiv M/\sqrt{2}$. Both complex scalar fields are charged under the same $U(1)$ symmetry, which is spontaneously broken through (\ref{pot}). The common phase $\phi$ of the two fields, $\phi_1 = M e^{i\phi/M}/\sqrt{2}$ and  $\phi_2 = M e^{-i\phi/M}/\sqrt{2}$,
is a gauge singlet boson that can be taken as the inflaton field\footnote{The phase $\phi$ being a
gauge singlet, one can assign an arbitrary potential for it. This is different from axion-like (pseudo-Goldstone boson) potentials that are constrained by a shift symmetry.}.
The remaining field content in eq.~(\ref{WLI}),
$\sigma$ is a scalar singlet,  $\psi_{\sigma R}$ are chiral fermions carrying the
same charge of $\psi_{1}$ and $\psi_{2}$, while $\psi_{\sigma L}$ has
zero charge and $\chi_1$ and $\chi_2$ are complex scalar fields.  The model satisfies the 
combined $Z_4$ and mirror symmetries\footnote{These symmetries are reminiscent of the ones also seen in the little Higgs and twin Higgs type of models~\cite{Schmaltz:2005ky,Chacko:2005pe}. {}For other recent connections of WI with standard model interactions and beyond, see, for example refs.~\cite{Berghaus:2025dqi,ORamos:2025uqs,Ito:2025lcg,Chakraborty:2025yms}.}
$\phi_1\leftrightarrow i\phi_2$, $\psi_{1L,R}\leftrightarrow \psi_{2L,R}$ and $\chi_1 \to \chi_2$, which protects the
inflaton potential to receive any large quantum or thermal corrections (see, e.g.~\cite{Bastero-Gil:2016qru,Bastero-Gil:2019gao} for details).

When the Yukawa couling $g_{\phi\psi}$ is dominant over the biquadratic $g_{\phi\chi}$ one,
we denote the model as the WLI fermionic (WLIF) model, while in the opposite case, 
with the biquadratic interactions between the $\phi_i$ and $\chi_i$ fields dominating over the
Yukawa one between $\phi_i$ and $\psi_i$, we denote the model as the WLI scalar (WLIS) model.
The dissipation coefficient evaluated in the WLIF case, in the high-temperature regime $g_{\phi\psi}M/T \ll 1$, and found to be given by~\cite{Bastero-Gil:2016qru}
\begin{eqnarray}
\Upsilon_{WLIF} = \alpha (h_{\sigma\psi}) \frac{g_{\phi\psi}^2}{h_{\sigma\psi}^2} \,T,\quad g_{\phi \chi}\to 0,
\label{UpsWLIF}
\end{eqnarray}
where $\alpha(h_{\sigma\psi})\simeq 3/[ 1-0.34\ln (h_{\sigma\psi})]$.
In the case of the WLIS model, the dissipation
coefficient was found to be given by~\cite{Bastero-Gil:2019gao}
\begin{equation}
\Upsilon_{WLIS}\simeq {4 g_{\phi\chi}^4\over h_{\chi\psi}^2}{M^2 T^2\over
  \tilde{m}_\chi^3}\left[1+{1\over\sqrt{2\pi}}\left({\tilde{m}_\chi\over
    T}\right)^{3/2}\right]e^{-\tilde{m}_\chi/T},\quad g_{\phi \psi} \to 0,
\label{UpsWLIS}
\end{equation}
where $\tilde{m}_\chi^2 \simeq g_{\phi\chi}^2M^2/2+\alpha^2T^2$, with $\alpha^2\simeq
\left[h_{\chi\psi}^2+ \lambda_{\chi_i\chi_j} \right]/12$, where $\lambda_{\chi_i\chi_j}$
indicates possible inter- and intra-interactions among the complex scalars $\chi_i$ and that are allowed under the symmetries of the model.

In the WLIF model, the form of the dissipation coefficient leads to  a growing function $G(Q)$ in the power spectrum, which in general drives the tilt
faster to a blue value ($n_s >1$). 
On the other hand, in the WLIS model, the dissipation coefficient behaves with a negative power in the temperature,
$\Upsilon_{WLIS} \propto 1/T$ for appropriate choices of model parameters. In this case, the function $G(Q)$ behaves as a decreasing function with increasing $Q$.
These two types of behavior that appear depending on the temperature dependence of the dissipation coefficient will be explicitly verified when we analyze the results from the full model with the interactions (\ref{WLI}), with the dissipation coefficient given by
\begin{equation}
\Upsilon = \Upsilon_{WLIF}+\Upsilon_{WLIS}.
\label{Upsilon}
\end{equation}

In the next section, we present our results using eq,~(\ref{Upsilon}) and contrast them with the recent P-ACT-LB-BK18 data and also with the earlier Planck data.

\section{Results}
\label{results}

An analysis of the individual separate dissipation coefficients (\ref{UpsWLIF}) and (\ref{UpsWLIS}) was done earlier in refs.~\cite{Bastero-Gil:2016qru} and \cite{Bastero-Gil:2019gao}, respectively. However, a unified analysis of the combined effect of both is still missing. We intend to fill this lacuna here. 

We present the results considering the total dissipation coefficient
eq.~(\ref{Upsilon}) for the WLI model eq.~(\ref{WLI}).
Combining eqs.~(\ref{UpsWLIF}) and (\ref{UpsWLIS}), we can write
\begin{eqnarray}
\Upsilon &=&  \frac{g_{\phi\psi}^2}{h_{\sigma\psi}^2} \left\{ \frac{3}{1-0.34 \ln (h_{\sigma\psi})}\,T \right.
\nonumber \\
&+& \left. b \,\frac{\left(2g_{\phi\chi} M \, T\right)^2 }{
  \tilde{m}_\chi^3}\left[1+{1\over\sqrt{2\pi}}\left({\tilde{m}_\chi\over
    T}\right)^{3/2}\right]e^{-\tilde{m}_\chi/T}\right\},
    \nonumber \\
\label{UpsWLIFS}
\end{eqnarray}
where $b$ is the ratio of the coupling constants,
\begin{equation}
b=  \frac{g_{\phi\chi}^2}{g_{\phi\psi}^2}  \frac{h_{\sigma\psi}^2}{h_{\chi\psi}^2},
\label{bvalue}
\end{equation}
and parameterizes the relative magnitude between the dissipation coefficients (\ref{UpsWLIF}) and (\ref{UpsWLIS}).

As already emphasized in the previous two sections, the strong dissipation regime of WI, where $Q>1$, is of particular interest. Hence, this is a regime that we will seek here. It will also guide us in the selection of the many possible parameters choices that can be made in the model~(\ref{WLI}). {}For instance, in the WLIF model, which was first studied in~\cite{Bastero-Gil:2016qru}, the form of the dissipation coefficient (\ref{UpsWLIF}) leads to a growing function $G(Q)$ in the power spectrum (\ref{full-power}) and which in general drives the spectral tilt faster to a blue value ($n_s >1$) when $Q$ to goes to values above one. The simplest inflaton potential that can be considered in this case and that leads to values for the tensor-to-scalar ratio $r$ and tilt $n_s$ consistent with observations is a quartic monomial potential. However, the consistency with the observations favors the weak dissipation regime
of WI, $Q <1$. Now in the case of the WLIS model, the dissipation coefficient (\ref{UpsWLIS}) behaves mostly of the background dynamics with a negative power in the temperature,
$\Upsilon \propto 1/T$ for appropriate choices of the model parameters. In this case, the function $G(Q)$ can display a decreasing function with increasing $Q$. In this regime of parameters, we can more easily have a strong dissipation with consistent values for $r$ and $n_s$. {}Furthermore, we can simply work with a quadratic monomial potential for the inflaton,
\begin{equation}
V(\phi) = \frac{m_\phi^2}{2} \phi^2.
\label{pot2}
\end{equation}

\subsection{Updated results for the WLIF and WLIS models}

Before analyzing the complete WLI model with the combined dissipation coefficient (\ref{UpsWLIFS}), it is useful to revise the two cases
WLIF and WLIS separately. The purpose for this is two-fold. Firstly, because they represent limiting cases for the complete model and will act as sanity checks of our main result. Secondly, because the analysis performed in the earlier references made use of simplified 
semi-analytical expressions to derive the observable quantities like $r$ and $n_s$, hence lacking precision as required when comparing with the most state of the art cosmological data. Here, this comparison between the predictions from the theory and observations is facilitated with the use of the recently released \texttt{WI2easy} code~\cite{Rodrigues:2025neh}. \texttt{WI2easy} is a dedicated code for WI dynamics, enabling efficient computation of both background evolution and curvature perturbations that makes possible precision comparisons between WI predictions and observational constraints.

\begin{center}
\begin{figure}[!bth]
\centerline{\subfigure[WLIF]{\includegraphics[width=7.5cm]{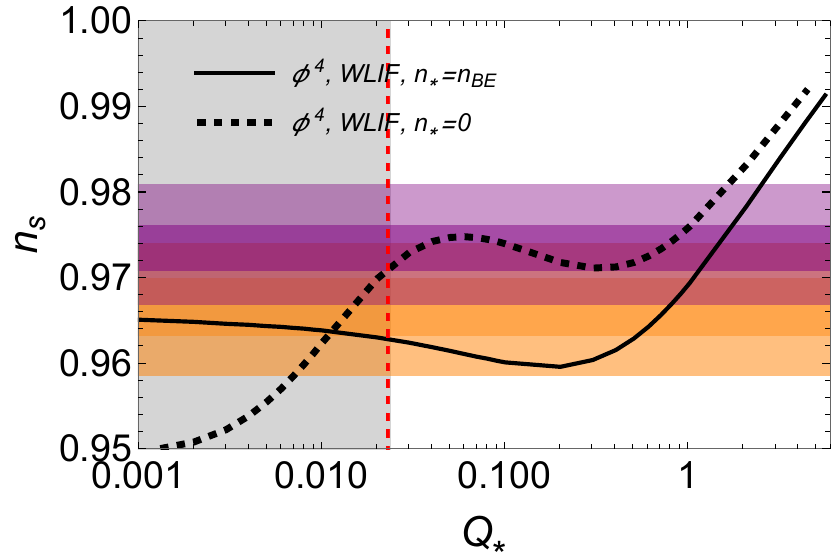}}
\subfigure[WLIS]{\includegraphics[width=7.5cm]{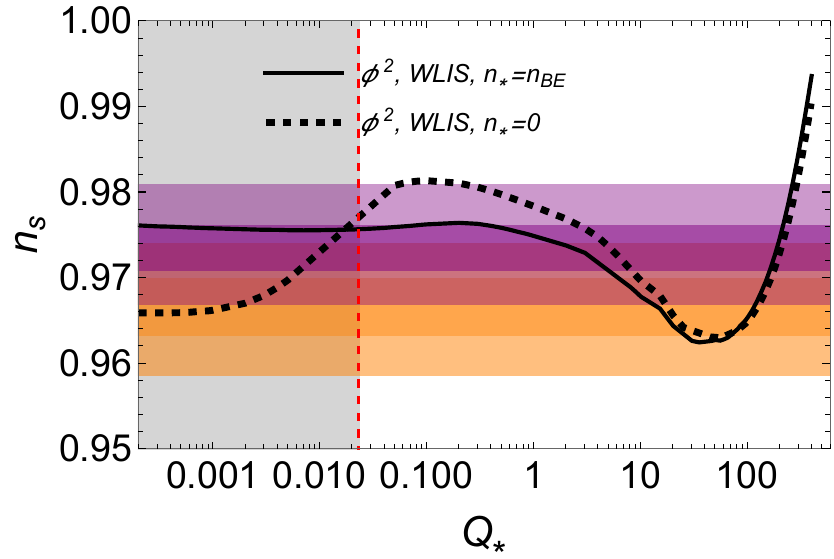}}}
\centerline{\subfigure[WLIF]{\includegraphics[width=7.5cm]{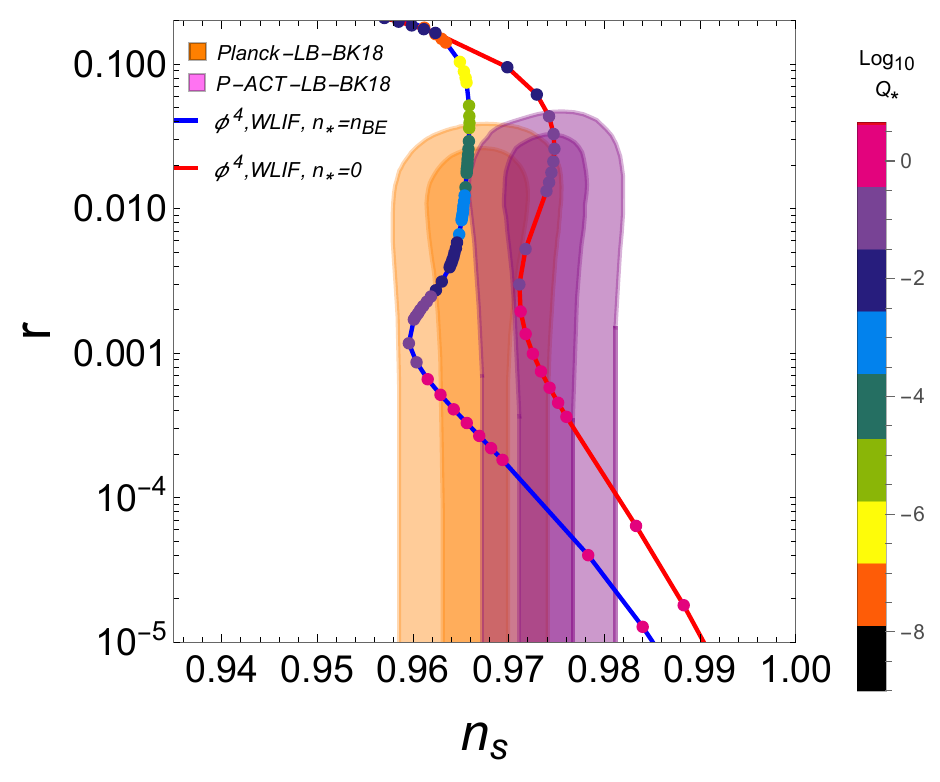}}
\subfigure[WLIS]{\includegraphics[width=7.5cm]{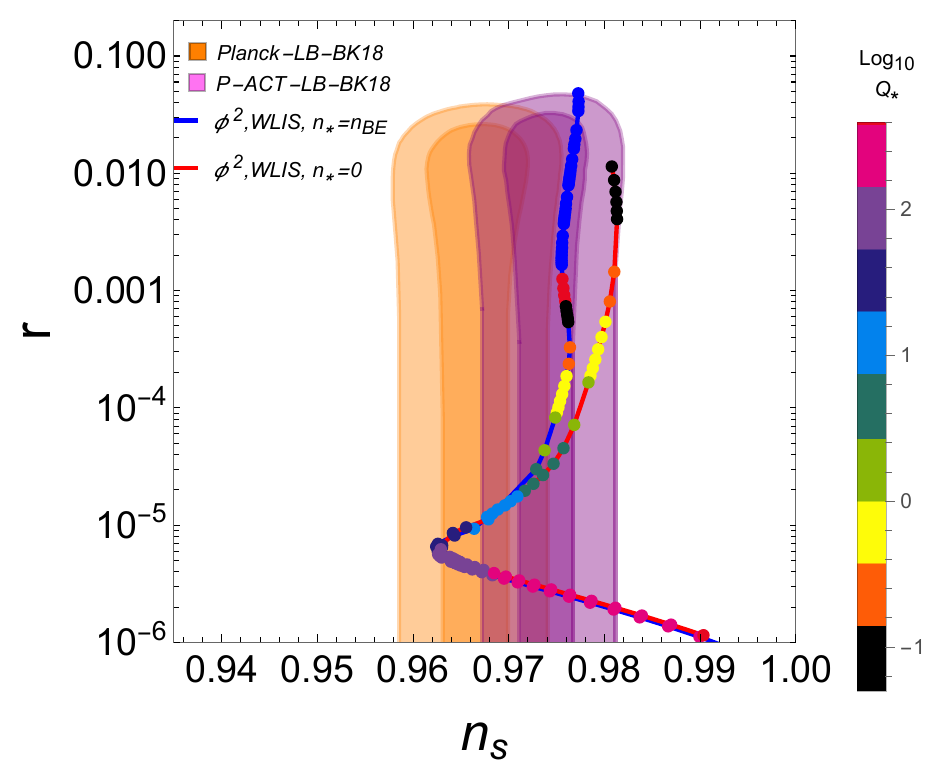}}}
\caption{The one-sigma and two-sigma constraints from Planck (orange shaded regions) and the combined P-ACT-LB results (violet shaded region) 
for $n_s$ as a function of $Q_*$ (panels a, b) and for $n_s \times r$ (panels c, d). Panels a and c are the results for the WLIF models (with a quartic inflaton potential), while panels b and d are for the WLIS model (with a quadratic inflaton potential). The shaded region on the left of panels a and b indicates the region for which $r>0.036$.}
\label{fig1}
\end{figure}
\end{center}

In fig.~\ref{fig1}, we show the results for the two models, the WLIF and WLIS, obtained for the spectral tilt $n_s$ and for the
tensor-to-scalar ratio $r$ as the value of $Q_*$ is varied, with all quantities evaluated at the Hubble radius crossing point $N_*$ for the pivot scale (see, e.g. ref.~\cite{Rodrigues:2025neh} for details). Note that varying $Q_*$ corresponds to changes
in the parameters and couplings of the two  models.
To enable comparison with the earlier references on these models, to obtain the results shown in fig.~\ref{fig1} the parameters considered in each case were chosen to be the same as those used in~\cite{Bastero-Gil:2016qru} (for the WLIF model) and~\cite{Bastero-Gil:2019gao}  (for the WLIS model) and also with the choices of a quartic inflaton potential $V=\lambda \phi^4/4$ for the WLIF model and a quadratic potential, eq.~(\ref{pot2}), for the WLIS model. Note that the WLIF model does not lead to satisfactory results, when compared with observations, when a quadratic potential is used (see also the results for the combined dissipation shown below)\footnote{See also ref.~\cite{Ballesteros:2023dno}, where this case of a quadratic potential with a dissipation coefficient that has a linear dependence on the temperature was recently analyzed.}.

The behavior seen in the curves for the WI models in fig.~\ref{fig1} in $r \times n_s$ is a
consequence of the relative
importance of each term in the power spectrum 
eq.~(\ref{full-power}) and how
$Q$ and $T/H$ evolve (see, e.g. ref.~\cite{Das:2020lut} for a discussion of the dynamical evolution for 
$Q$ and $T/H$ in WI and for their effect on $n_s$, see ref.~\cite{Das:2022ubr}).

The results in the fig.~\ref{fig1} show that
while the WLIF model produces results compatible with both Planck and ACT data mostly in the weak dissipative regime ($Q \lesssim 1$), the WLIS model can easily reach the strong dissipative regime,
having a wide range of parameter values where
$Q \gg 1$ covers the data regions.
In particular, in the WLIS model $Q$ can be sufficiently large such that the inflaton excursion is sub-Planckian, $\Delta \phi \lesssim M_{\rm Pl}$. This happens when $Q \gtrsim 100$. The values of $Q$ beyond which we have $\Delta \phi < M_{\rm Pl}$ are marked in {}Fig.~\ref{fig1}(b) by the stars. In particular, we can consider the value for the dissipation ratio as $Q_*=175$, for which we obtain $n_s=0.9708$ and $r=3.41\times 10^{-6}$, when $n_*=0$ and  $n_s=0.9722$ and $r=3.03\times 10^{-6}$, when $n_*=n_{\rm BE}$, hence both cases lying around the 68$\%$ C.L. region for both Planck and ACT. {}For both cases we have $\Delta \phi \simeq 0.7 M_{\rm Pl}$ while also having $m_\phi \simeq 6 \times 10^{-7}M_{\rm Pl}$
and $H_*\simeq 1.8 \times 10^{-7} M_{\rm Pl}$, hence $m_\phi > H_*$, thus evading
possible $\eta$-problems. In addition, we also have $\epsilon_V \simeq \eta_V \simeq 3.6$, which combined with the sub-Planckian field excursion
and $m_\phi > H$ results, makes this WI model consistent with effective
field theories according to the swampland program~\cite{Motaharfar:2018zyb, Das:2018rpg,Brandenberger:2020oav, Berera:2019zdd,Berera:2020dvn}.

\begin{center}
\begin{figure}[!bth]
\subfigure[]{\includegraphics[width=7.5cm]{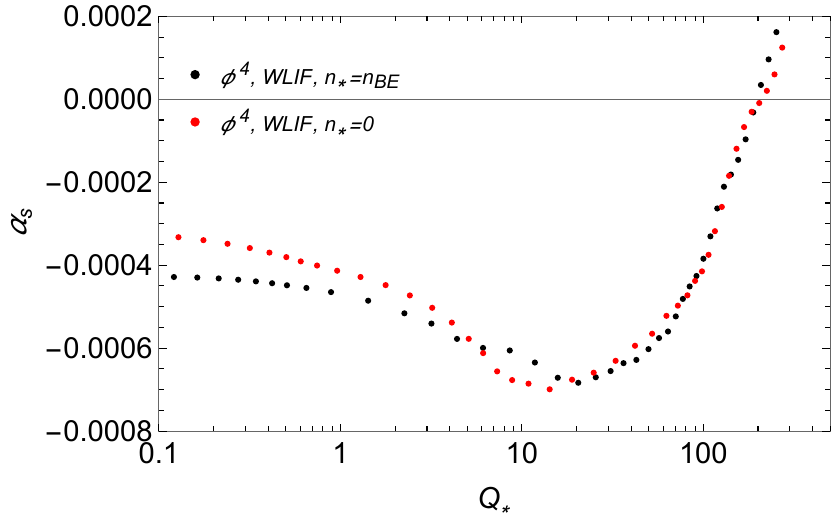}}
\subfigure[]{\includegraphics[width=7.5cm]{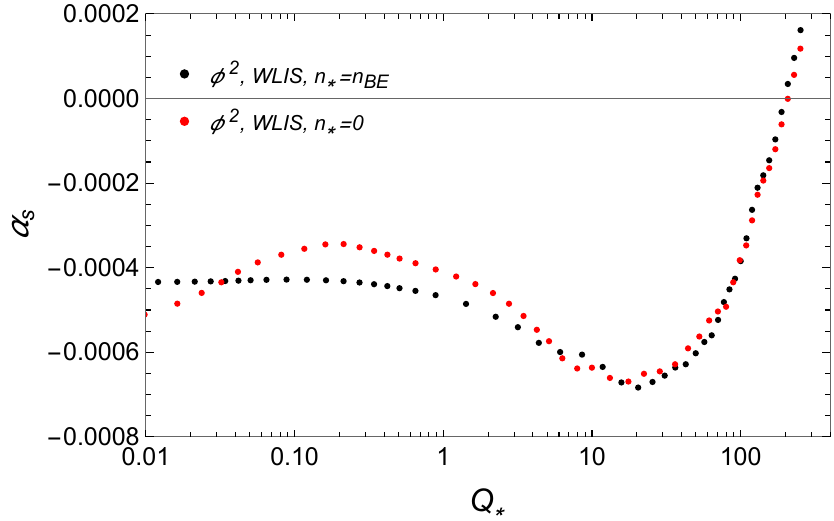}}
\caption{Similar to the {}Fig.~\ref{fig1}, but now showing the running $\alpha_s$ as a function of the dissipation ratio $Q_*$ for the WLIF and WLIS models.}
\label{fig2}
\end{figure}
\end{center}

The previous references for the WLIF and WLIS models have not analyzed the behavior for the running $\alpha_s$ of the spectral tilt. Since the recent
ACT data tend to slightly favor a positive running (though still compatible with a null value), it is also opportune to also make this analysis of how $\alpha_s$ behaves with $Q$ for the WLIF and WLIS models.
This result is shown in fig.~\ref{fig2}.

It is interesting to see
that WI can also easily accommodate positive values for $Q\gg 1$. The physical reason behind this is that the negative values for $\alpha_s$ in cold models arise from the fact that the inflationary trajectories deviate from quasi-dS at the end of inflation when the shortest wavelength modes exit the horizon. However, for WI the energy density of the inflaton continuously decreases, leading to a smooth transition to radiation domination (which in most cases is accompanied by an increase of $Q$). In {}Fig.~\ref{fig2},  the solid line corresponds to the results when neglecting $n_*$
in the equation for the scalar spectrum in WI or taking $n_*$ as a Bose-Einstein distribution (dashed line). The results fall well within the one-sigma
region found by ACT~\cite{ACT:2025tim} for $\alpha_s$. It also shows that for large $Q$, the running tends to move from negative (in the weak dissipation regime of WI) towards positive values (corresponding to the strong dissipative WI regime), while it still remains very small.  

\subsection{Results for the full WLI model}
\label{results-WLI}

Let us now turn to the presentation of the results for the full model
(\ref{WLI}), with dissipation coefficient given by eq.~(\ref{UpsWLIFS}). Once again, we seek parameters that allow for the strong dissipative regime of WI, while also using the simplest inflaton potential eq.~(\ref{pot2}). As a representative set of parameters, we use those that already allowed 
strong dissipation in the case of the WLIS model analyzed in fig.~\ref{fig1}(a). That is, with fixed value $g_{\phi\chi} M = 2.6 \times 10^{-5} M_{\rm Pl}$, coupling $\alpha^2= 1/8$ and $h_{\sigma\psi}=2$. The remaining parameters, for example $m_\phi$, the individual values for the coupling $g_{\phi\chi}$ and mass scale $M$,
along with the remaining coupling constants in (\ref{WLI}) are all self-consistently determined. In particular, $m_\phi$ is evaluated by requiring that the scalar of curvature power spectrum at the pivot scale $k_* = 0.05/Mpc$ has the CMB amplitude normalization~\cite{Aghanim:2018eyx} $A_s \simeq 2.105 \times 10^{-9}$. The coupling constants and the mass parameter $M$ can be determined for each value of $Q$ by associating them with the chosen value for the ratio $b$ defined in  
eq.~(\ref{bvalue}) and also with the overall prefactor $g_{\phi\psi}^2/h_{\sigma\psi}^2$ in eq.~(\ref{UpsWLIFS}), which in the 
\texttt{WI2easy} code, it is associated with the $C_\Upsilon$ constant defined in there.
We consider three representative cases for the ratio of coupling constants defined in eq.~(\ref{bvalue}), $b=1,\, 10,\, 20$.  

\begin{center}
\begin{figure}[!bth]
\centerline{\includegraphics[width=8.8cm]{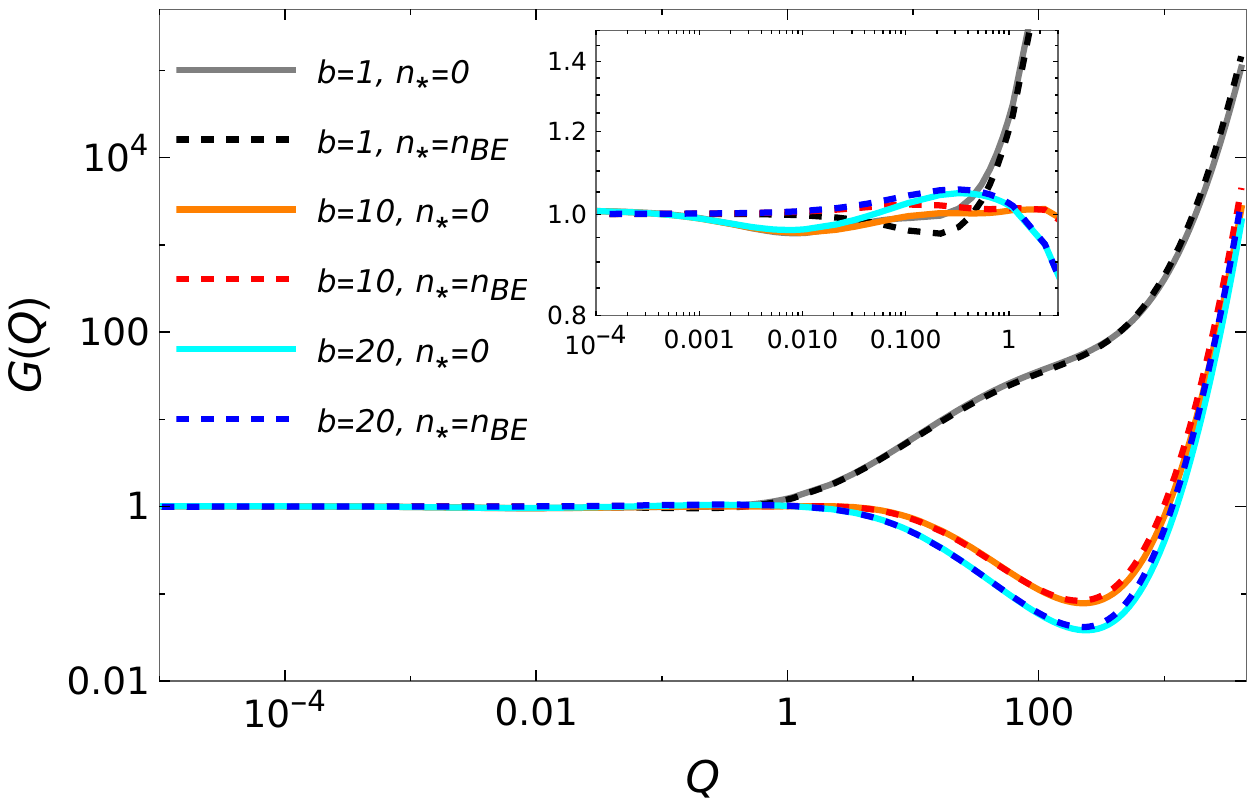}}
\caption{The $G(Q)$ function generated through  
\texttt{WI2easy}~\cite{Rodrigues:2025neh} for a dissipation coefficient eq.~(\ref{UpsWLIFS}) for $b=1,\, 10,\, 20$, and for the potential (\ref{pot2}). We present the result for both nonthermal $n_*=0$  and fully thermalized $n_*=n_{\rm BE}$. The inset plot zoom in around the region $10^{-4} \lesssim Q \lesssim  1$.}
\label{fig3}
\end{figure}
\end{center}

In fig.~\ref{fig3} we show the numerically generated results for $G(Q)$ through \texttt{WI2easy} and that is necessary in the expression for the power spectrum (\ref{full-power}). We consider both the cases of including or not the thermal distribution term $n_*$ in eq.~(\ref{Fk}). In appendix~\ref{appA} we give a justification of when this term
might contribute in the power spectrum in the case of the WLI model.
The results presented in fig.~\ref{fig3} show that there is no significant differences for $G(Q)$ for the present model when considering $n_*=0$  or $n_*=n_{\rm BE}$, except in the region $10^{-4} \lesssim Q \lesssim  1$. We note that the results for $G(Q)$ are only sensitive to the presence or not of $n_*$ in the weak dissipative regime. In obtaining the numerical results for $G(Q)$, we have also explicitly taken into account a stochastic noise term in the radiation perturbation equation that comes as a consequence of the conservation of the total stress energy tensor. This term was first discussed in \cite{Bastero-Gil:2014jsa} and it also affects the perturbations (through $G(Q)$) in the region $10^{-4} \lesssim Q \lesssim  1$. Earlier fittings that were proposed for $G(Q)$, e.g. in ref.~\cite{Bastero-Gil:2011rva} and also more recently in \cite{Montefalcone:2023pvh}, have not taken into account the possibility of having $n_*=0$ or the presence of the above stochastic term in the perturbation equation for the radiation. Those fitting results for $G(Q)$ should then be used with care. In all of our results presented here, including the ones in figs.~\ref{fig1} and \ref{fig2}, we have explicitly included the stochastic term in the perturbation equation for the radiation (see also 
ref.~\cite{Rodrigues:2025neh} for details).

\begin{center}
\begin{figure}[!bth]
\subfigure[$b=1$]{\includegraphics[width=5cm]{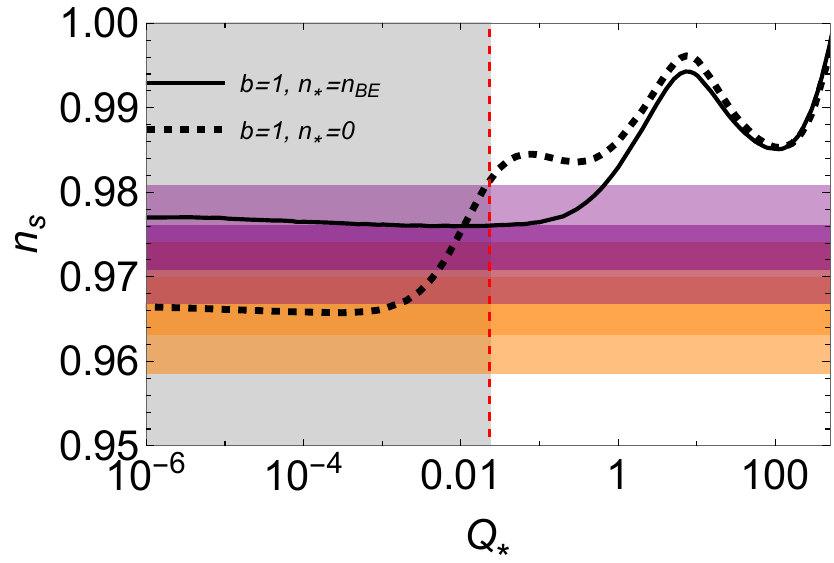}}
\subfigure[$b=10$]{\includegraphics[width=5cm]{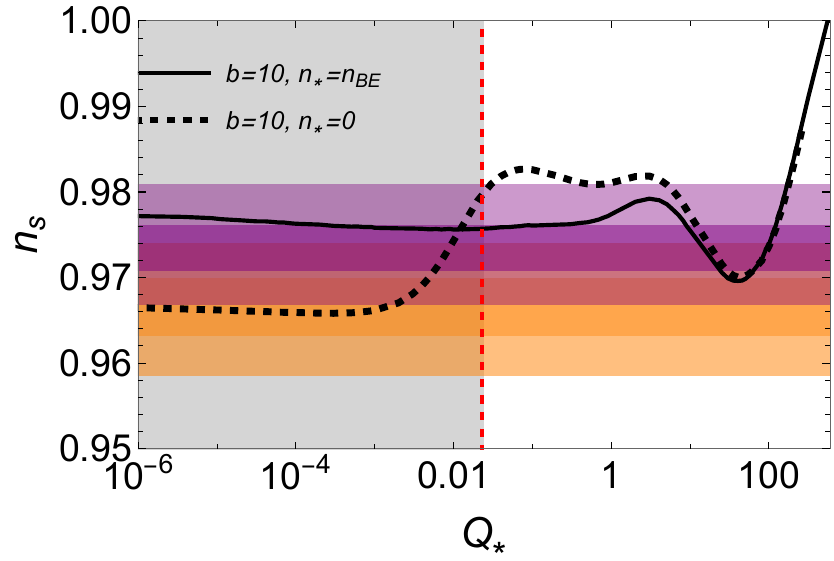}}
\subfigure[$b=20$]{\includegraphics[width=5cm]{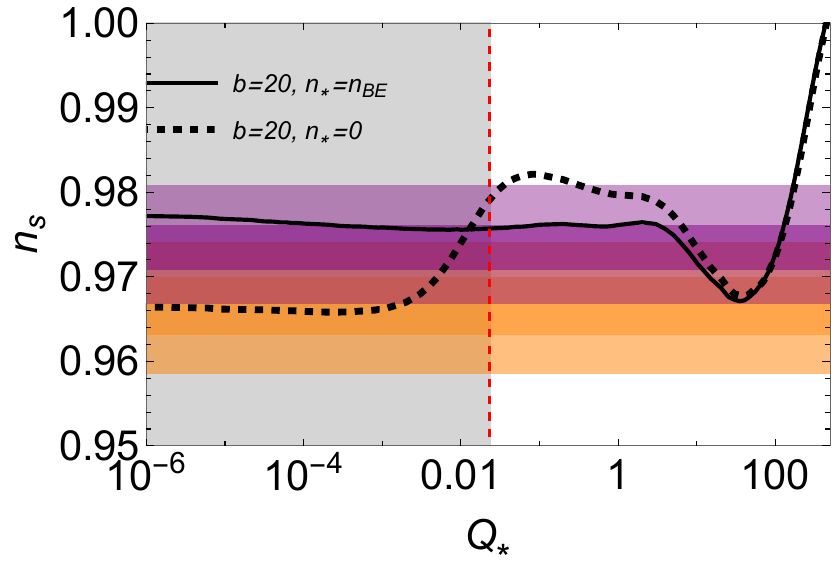}}
\subfigure[$b=1$]{\includegraphics[width=5cm]{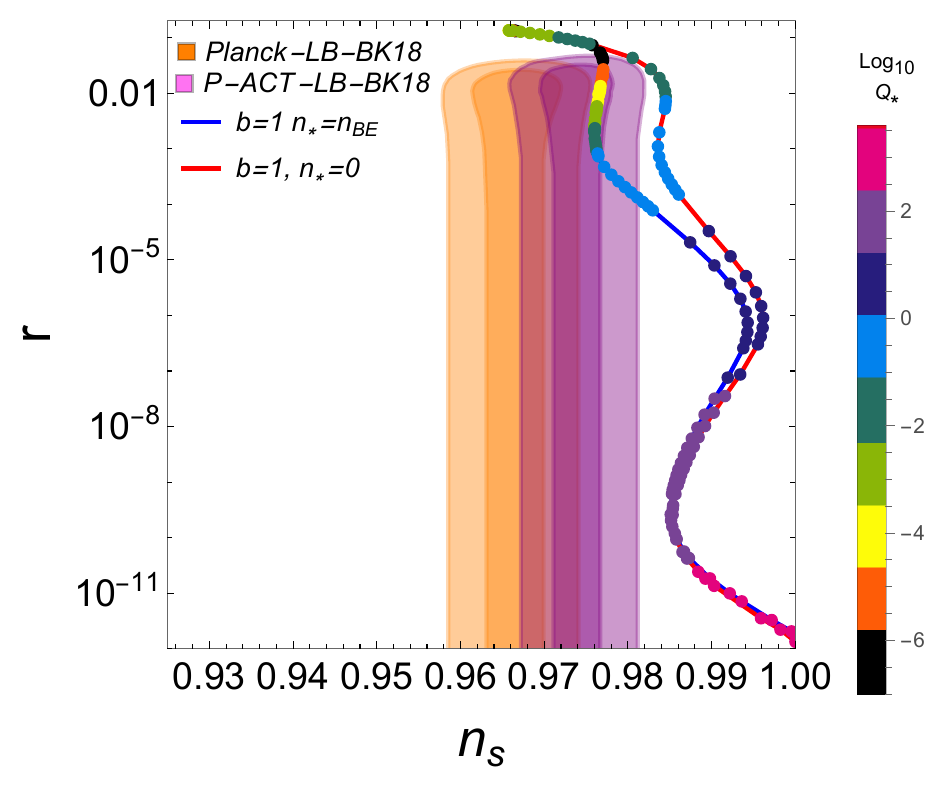}}
\subfigure[$b=10$]{\includegraphics[width=5cm]{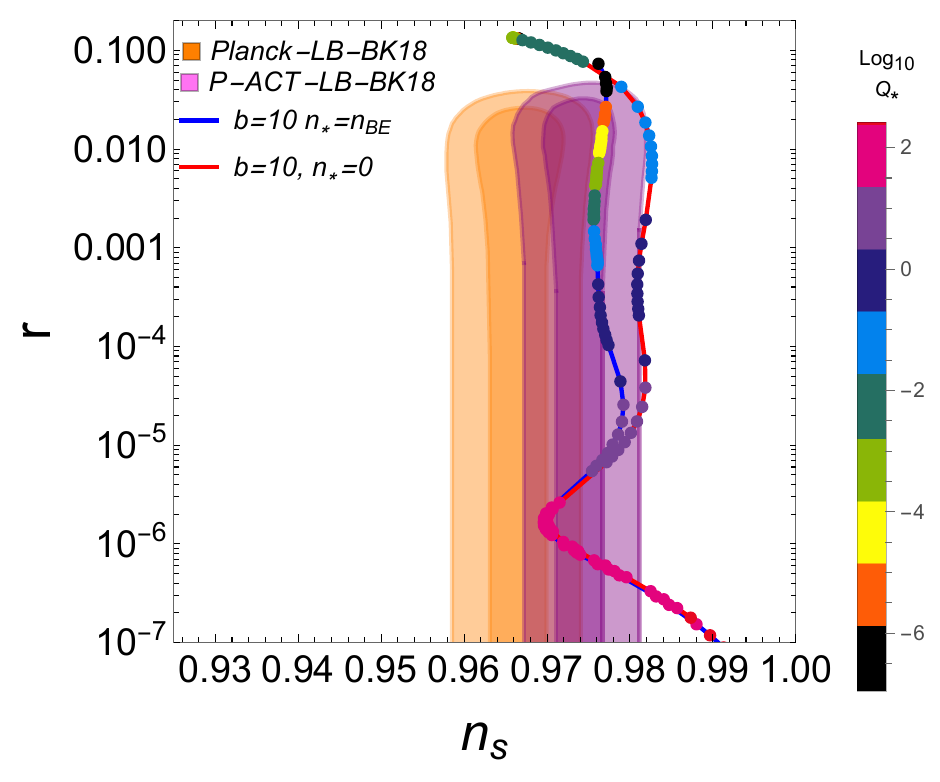}}
\subfigure[$b=20$]{\includegraphics[width=5cm]{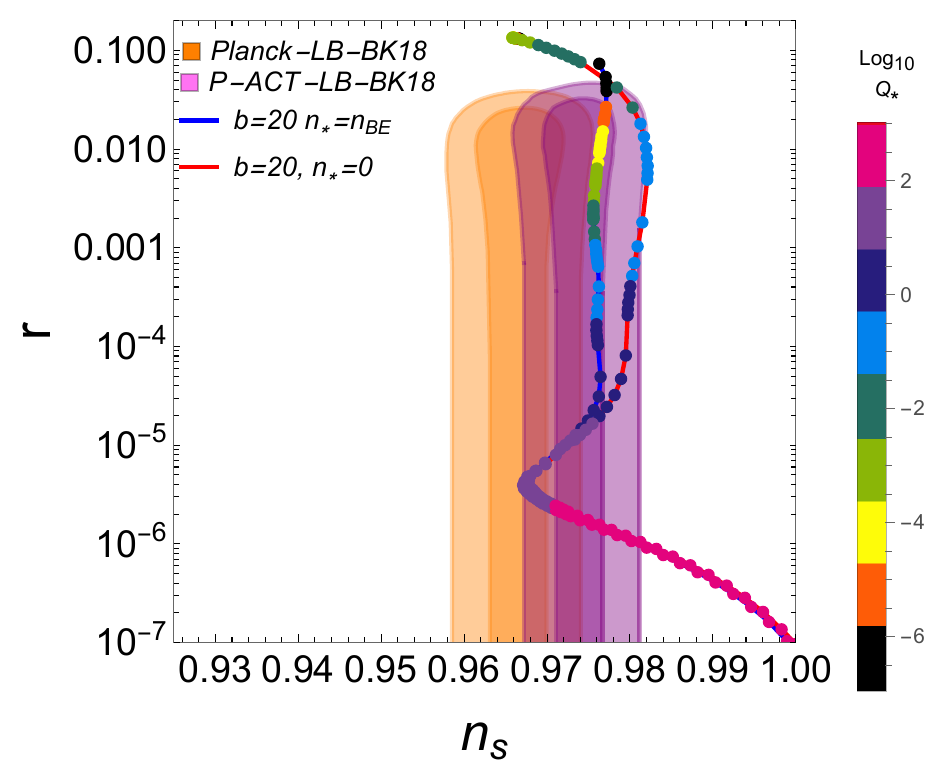}}
\caption{The same as in fig.~\ref{fig1}, but showing the results for the WLI model, with a quadratic inflaton potential and for the cases of $b=1,\,10$ and $20$. Panels a, b and c show $n_s$ as a function of $Q_*$, while panels d, e and f show $n_s \times r$.  }
\label{fig4}
\end{figure}
\end{center}

In fig.~\ref{fig4}, we present the results for $n_s$ and $r$ obtained for the WLI model, with the dissipation coefficient given by eq.~(\ref{UpsWLIFS}). We consider again the results when $b=1,\, 10,\, 20$. Other cases with $b < 1$ tend only to allow values
of $r$ and $n_s$ compatible with the observations when the inflaton is thermalized, i.e., $n_* = n_{\rm BE}$ in eq.~(\ref{Fk}), but even so this only happens in the very weak dissipative regime of WI, $Q \ll 1$.
As the value of $b$ is increased, the dissipation coefficient (\ref{UpsWLIS}) starts to become more dominant throughout the dynamics,
favoring the strong dissipative regime of WI, consistent with what is shown
in fig.~\ref{fig1} for the WLIS model.

\begin{center}
\begin{figure}[!bth]
\subfigure[$b=1$]{\includegraphics[width=5cm]{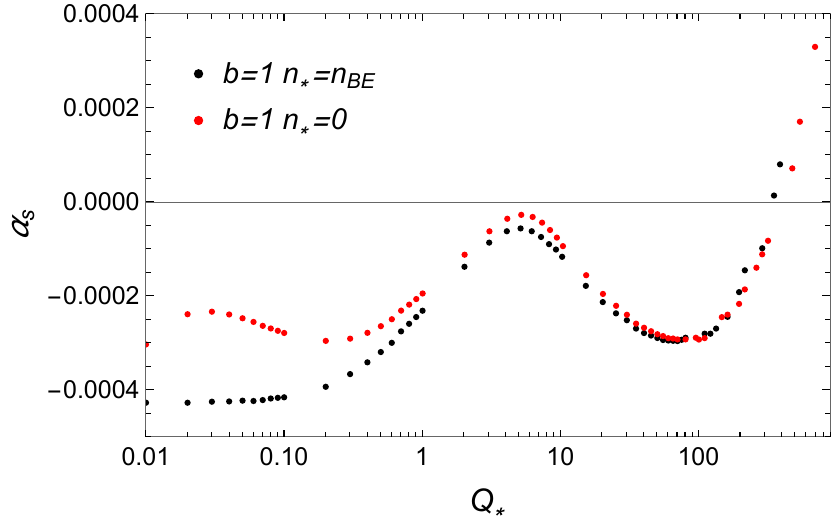}}
\subfigure[$b=10$]{\includegraphics[width=5cm]{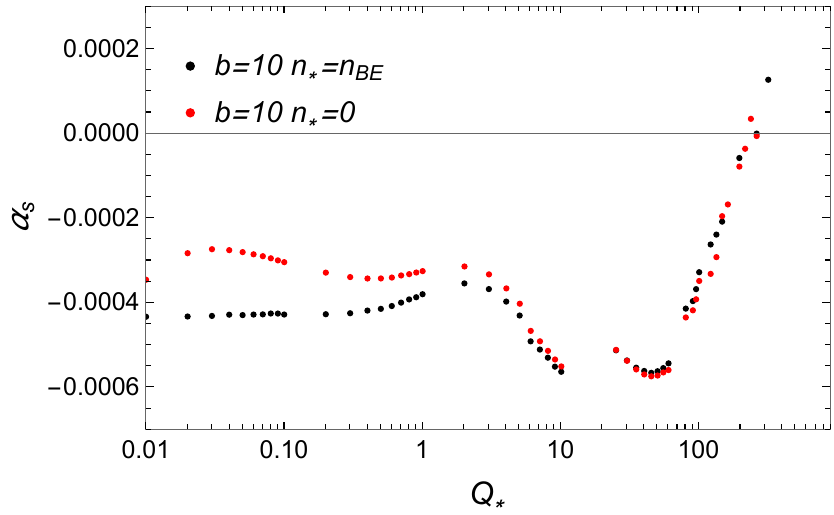}}
\subfigure[$b=20$]{\includegraphics[width=5cm]{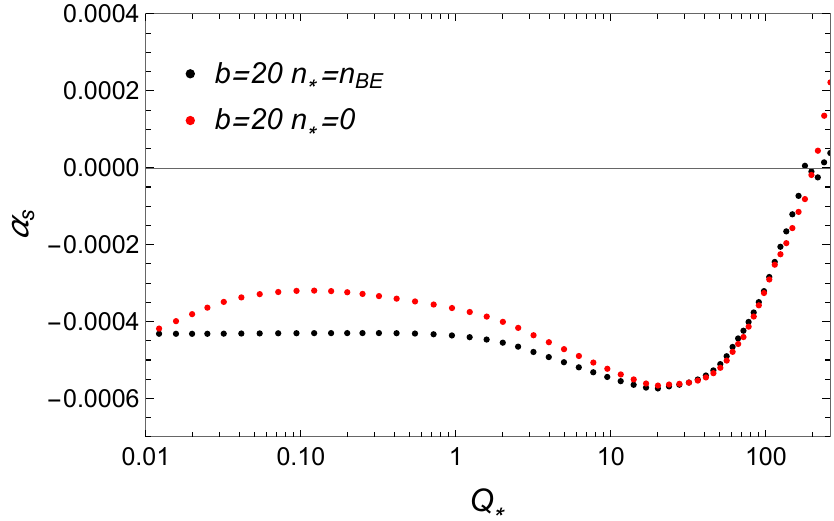}}
\caption{The running $\alpha_s$ for the WLI model as a function of the dissipation ratio $Q_*$ at Hubble crossing and with the dissipation coefficient (\ref{UpsWLIFS}). }
\label{fig5}
\end{figure}
\end{center}

{}Finally, in fig.~\ref{fig5}, we show the results for the running $\alpha_s$ for the WLI model and for the three cases of ratio of coupling constants $b$ considered above. We again see the trend observed for the separated cases of the WLIF and WLIS models, where the running tends to become positive for $Q \gtrsim 100$.

\section{Conclusions}
\label{conclusions}

In this paper, we have built upon earlier results obtained for the warm little inflaton model first  studied in ref.~\cite{Bastero-Gil:2016qru} when considering the leading dissipative effects coming from fermion interactions with the inflaton and then in ref.~\cite{Bastero-Gil:2019gao} for scalar interactions, by further generalizing the model. We have presented an analysis not done before for the combined use of fermionic and scalar interactions, in which case the dissipation coefficient in the model is extended to eq.~(\ref{UpsWLIFS}). Moreover, our results were computed using the recently released code \texttt{WI2easy}~\cite{Rodrigues:2025neh}, which provides the best precision analysis to date of background dynamics and cosmological perturbations in WI.  Representative sets of parameters for the model have been considered, for which we have evaluated the relevant inflationary observable quantities, the tensor-to-scalar ratio $r$, the spectral tilt $n_s$ and the running of the tilt $\alpha_s$. We then compared the results for these with both the earlier Planck data and the more recent ACT data. As a consequence of the ACT data showing a preference for slightly larger values for $n_s$ compared to earlier Planck data, our results show that this tends to favor larger values for the dissipation coefficient ratio $Q$ in WI for the model studied here.

Our results in general also point out to possible implications of the P-ACT-LB-BK18 data as far as inflationary model-building is concerned. At face value, a tiny $r$ and $n_s$ very close to unity tell us that the standard models of renormalizable, monomial potentials seem to fall out of favor of observations. However, the deeper revelation is that observations seem to strike at the heart of the problem of protecting the flatness of the potential from quantum corrections. One can always find one- or two-parameter extensions of current models of cold inflation which fit the data as more precise mappings of the CMB come in. Yet even accepting that, it has not been the same model that is relevant as the data changes, but rather the model has changed, selected from effectively an infinite space of models. Moreover, it is time we introspect to ask if such extensions have an embedding in
any UV-theory. Put another way, can one accentuate well-motivated models with such ad hoc couplings that are neither derived from any underlying theory nor protected from the effect of Planck-suppressed operators? Even when one takes the modern view of EFT that non-renormalizable
potentials are, by themselves, not problematic and simply point towards a UV-cutoff to which scale it can be trusted, it makes little sense to pick and choose between operators (of the same derivative order) such that a model becomes viable. This just turns model building into a speculative exercise rather than based on a theoretically reasoned foundation \cite{Berera:2023liv}. A principled path forward for inflation seems to require a new physical input which would help reconciling theoretically-consistent models with observations.

In this paper, we argue that WI is the right paradigm to do so. It is only natural that the inflaton field interacts with other fields during inflation and the only further assumption WI makes is regarding the strength of the dissipative terms generated in the process. However, one  then finds that standard models of WI can easily fit the data where their cold counterparts failed. One might question why the specific potential chosen in this paper is not equally fine-tuned. Our main argument is that the exact choice of potential is not as important in WI as it is in the cold case. This is precisely because of the presence of the extra dissipative terms. We could just as easily have chosen some other potential and much of our discussion would still go through. Moreover, we do choose a form of the potential which is renormalizable and thereby can be assumed to be protected from Planckian corrections. As can be clearly seen from {}Fig.~\ref{fig1}, dissipative corrections make even these monomial potentials viable in WI. Note that although the strong dissipative regime would be preferred from theoretical considerations, these models produce the right spectrum even for small $Q$. Indeed, the results presented here simply confirm that strong dissipative regimes (of warm little models) are preferred by data and are not obtained by any fine-tuning to fit new observations. However, future observations of the tensor-to-scalar ratio, or the lack thereof, can be used as a signature of the strong \textit{vs.} weak dissipative regime for WI. 

Naturally, WI is also not free from any constraints. Firstly, the dissipative coefficients need to be computed from first-principles and that is what has been done for WLI model considered here. More needs to be done regarding the microphysics of such dissipative interactions and how they can be derived from a full thermal field theory perspective in an accelerating background. (A possible application of this is the question of whether the inflaton thermalizes or not, something that we have considered, at least qualitatively in the appendix.) This can put further bounds on the parameter space of WI. Moreover, $Q$ is not a completely unconstrained parameter, even from observations. It is natural to expect that going to arbitrarily large values of the dissipative coefficient would lead to an unacceptably large $f_{NL}$ for the warm shape in the scalar bispectrum (we have made sure that going up to $Q\sim \mathcal{O}(100)$ is consistent with current bounds on the bispectrum \cite{Bastero-Gil:2019gao}). We plan to find the upper bound on $Q$ coming from the bispectrum, which would also restrict how large $n_s$ can be in WI. 

For completeness, we also mention the possibility that inflation itself might not be the right way to think about the early universe. Some alternatives to inflation \cite{Brandenberger:1988aj, Steinhardt:2001st, Agrawal:2020xek, Brahma:2022dsd, Boyle:2022lcq}, which are grounded in a fundamental UV-theory, might be better suited to explain observations. Interestingly, many of them also postulate thermal fluctuations \cite{Nayeri:2005ck, Brahma:2021tkh} as in WI, thereby pointing toward a larger application of dissipative corrections that go beyond inflation.

\acknowledgments
AB is partially funded by STFC. SB is supported in part by the Higgs Fellowship and by the STFC Consolidated Grant ``Particle Physics at the Higgs Centre''. 
R.O.R. acknowledges financial support by research grants from Conselho
Nacional de Desenvolvimento Cient\'{\i}fico e Tecnol\'ogico (CNPq),
Grant No. 307286/2021-5, and from Funda\c{c}\~ao Carlos Chagas Filho
de Amparo \`a Pesquisa do Estado do Rio de Janeiro (FAPERJ), Grant
No. E-26/201.150/2021. 
The work of G.S.R. is supported by a PhD scholarship from FAPERJ.
{}For the purpose of open access, the author has applied a Creative Commons
Attribution (CC BY) licence to any Author Accepted Manuscript version
arising from this submission.

\appendix
\section{Thermalization of the inflaton in the WLI model}
\label{appA}

Here, we provide estimates to justify whether the inflaton could thermalize in the WLI model, that is, whether $n_*$ in eq.~(\ref{Fk}) can assume a
Bose-Einstein distribution, $n_* \equiv n_{\rm BE}$.
The answer for this question is, of course, strongly dependent on the details of the microphysics of the specific WI model under study. This would involve, for example, whether scattering rates of the inflaton
with other fields could happen at a rate $\Gamma_{\rm scatt}$ fast enough compared to the expansion rate, $\Gamma_{\rm scatt} > H$. The condition $\Gamma_{\rm scatt} > H$ is usually assumed to justify thermalization of radiation particles in cosmology~\cite{Kolb:1990vq}, although a full justification can only be appropriately given through either an appropriate lattice simulation of the full dynamics,
or a dedicated kinetic Botzmann equation study for the occupation number for the inflaton in WI. A previous general analysis of the latter have first been given in ref.~\cite{Bastero-Gil:2017yzb}. 
The results obtained there in fact indicated that thermalization, in the sense that the distribution function can assume a thermal equilibrium form, is justified when the rate $\Gamma$ is of order or greater than the expansion rate $H$\footnote{In fact, the results in ref.~\cite{Bastero-Gil:2017yzb} have shown that even when $\Gamma < H$ and the system is out-of-equilibrium, the distribution function can assume a form proportional to the thermal one, with a prefactor given by $A \Gamma/(3H +\Gamma)$, where $A$ is a numerical coefficient satisfying $A \sim {\cal O}(1)$.}. In the absence of a more complete analysis for the thermalization of the inflaton, it is common in most applications in WI to
typically consider both possibilities --- non thermalized inflaton perturbations, $n_*=0$, or fully
thermalized, $n_*=n_{\rm BE}$ --- in the expression for the scalar of curvature power in WI, eq.~(\ref{full-power}). When presenting all our results, we have assumed the same philosophy. However, we can take advantage of the fact that, since we know the details of the microphysical WLI model, we know all relevant interactions involving the inflaton field eq.~(\ref{WLI}) and hence can estimate the relevant scattering rates that  contribute to the thermalization of the inflaton. 

In our estimates below, we will focus on the direct interactions of the inflaton with the light fields $\chi_{1,2}$ and $\psi_{1.2}$. We will also consider, for simplicity, only the $2\to 2$ processes that come from these interactions. {}From eq.~(\ref{WLI}), we have that the direct interactions of the inflaton field with $\chi_{1,2}$ and $\psi_{1.2}$ are
\begin{eqnarray} 
\mathcal{L}_{\rm int} &=& -g_{\phi\psi} M \cos(\phi/M) \bar\psi_{1L} \psi_{1R} -g_{\phi\psi} M \sin(\phi/M) \bar\psi_{2L} \psi_{2R}
\nonumber \\
&-& g_{\phi\chi}^2 M^2 \cos^2(\phi/M) |\chi_1|^2 -g_{\phi\chi}^2 M^2 \sin^2(\phi/M) |\chi_2|^2.
\label{Lscatt}
\end{eqnarray} 
When shifting the inflaton field around its background value and perturbations, $\phi \to \phi + \delta \phi$, we find that the relevant vertices for the processes involving the inflaton perturbations up to second order in $\delta \phi$ are
\begin{eqnarray} 
\mathcal{L}_{\rm int}(\delta \phi, \chi_i, \psi_i, \bar  \psi_i) &=& g_{\phi\psi} \sin(\phi/M) \delta \phi \bar\psi_{1L} \psi_{1R} -g_{\phi\psi} \cos(\phi/M) \delta \phi \bar\psi_{2L} \psi_{2R}
\nonumber \\
&+& g_{\phi\chi}^2 M \sin(2\phi/M) \delta \phi \left(|\chi_1|^2 - |\chi_2|^2\right)
\nonumber \\
&+& \frac{g_{\phi\psi}}{2 M} \cos(\phi/M) \delta \phi^2 \bar\psi_{1L} \psi_{1R} +\frac{g_{\phi\psi}}{2 M} \sin(\phi/M) \delta \phi^2 \bar\psi_{2L} \psi_{2R}
\nonumber \\
&+& g_{\phi\chi}^2 \cos(2\phi/M) \delta \phi^2 \left( |\chi_1|^2 -|\chi_2|^2 \right).
\label{Lscatt2}
\end{eqnarray} 
{}From the interaction Lagrangian density eq.~(\ref{Lscatt2}) we read off the vertices:
\begin{align}
&{\rm Yukawa\; (linear\; in\;} \delta\phi):\qquad
{\cal L}\supset y_1\,\delta\phi\,\bar\psi_{1L}\psi_{1R}-y_2\,\delta\phi\,\bar\psi_{2L}\psi_{2R},
\nonumber\\[4pt]
&{\rm trilinear\; (}\delta\phi\chi\chi\!):\qquad
{\cal L}\supset \kappa_\chi\,\delta\phi\,\left(|\chi_1|^2 -|\chi_2|^2\right),
\nonumber\\[4pt]
&{\rm contact\; (}\delta\phi^2\bar\psi\psi\!):\qquad
{\cal L}\supset \lambda_{\psi,1}\,\delta\phi^2\bar\psi_{1L}\psi_{1R}
+ \lambda_{\psi,2}\,\delta\phi^2\bar\psi_{2L}\psi_{2R},
\nonumber\\[4pt]
&{\rm contact\; (}\delta\phi^2\chi^2\!):\qquad
{\cal L}\supset \Lambda_\chi\,\delta\phi^2\left(|\chi_1|^2-|\chi_2|^2\right),
\label{eq:effint}
\end{align}
where we have defined the couplings
\begin{align}
&y_1\equiv g_{\phi\psi} S,\; y_2\equiv g_{\phi\psi} C,
\nonumber\\[4pt]
&\kappa_\chi\equiv g_{\phi\chi}^2 M S_2,
\nonumber\\[4pt]
&\lambda_{\psi,1}\equiv \frac{g_{\phi\psi}}{2M}C,\; \lambda_{\psi,2}\equiv \frac{g_{\phi\psi}}{2M}S,
\nonumber\\[4pt]
&\Lambda_\chi\equiv g_{\phi\chi}^2 C_2,
\label{eq:effcouplings}
\end{align}
and where $C,\, S,\, C_2$ and $S_2$ are defined as $S\equiv \sin(\phi/M),\, C\equiv \cos(\phi/M),\,
S_2\equiv\sin(2\phi/M),\, C_2\equiv\cos(2\phi/M)$, with $\phi$ denoting the background inflaton field.

\begin{center}
\begin{figure}[!bth]
\centerline{\includegraphics[width=12cm]{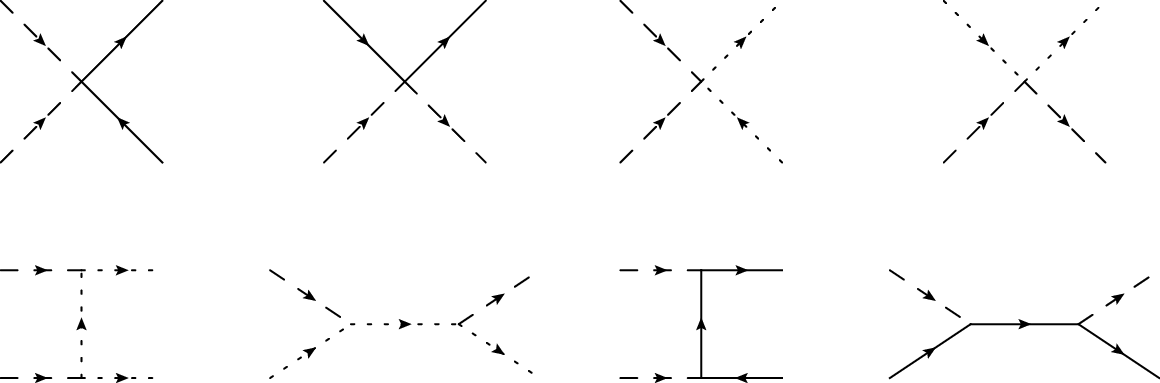}}
\caption{Feynman diagrams for the tree-level $2 \to 2$ scattering processes originating from 
the interaction terms in eq.~(\ref{Lscatt2}). The top diagrams come from contact interactions, while the bottom diagrams come from the trilinear vertex involving the scalars $\delta \phi$ and $\chi_i$ and from the Yukawa vertex between $\delta \phi$ and $\psi_i,\, \bar \psi_i$.
A dashed line represents the $\delta \phi$ excitation, a dotted line represents the
complex scalars $\chi_i$ and the solid lines the fermions $\psi_i$.}
\label{fig6}
\end{figure}
\end{center}

Examples of tree-level $2 \to 2$ scattering processes originating from 
the interaction terms in eq.~(\ref{eq:effint}) are shown in fig.~\ref{fig6}. Although a detailed computation of all thermal scattering rates that can contribute to the possible thermalization of the inflaton excitation $\delta \phi$ is beyond the scope of this paper, here we estimate the rates for some of the main processes leading to
thermalization of the inflaton particle states.

Considering an arbitrary $2\rightarrow2$ process involving relativistic particles, we have that the differential cross-section (for the $s$-channel) is defined as~\cite{Peskin:1995ev}:
\begin{equation} \label{differential_cross_section}
{d\sigma\over d\Omega}={|\mathcal{M}|^2\over 64\pi^2 s}~,
\end{equation}
where $\sqrt{s}$ is the centre-of-mass energy. We can make estimates of the magnitude of the different processes shown in fig.~\ref{fig6} by neglecting the angular correlations and we can take typical momenta and energies, both for external and internal legs, to be of the order of the temperature of the thermal bath, where we also assume that all particles are relativistic, with $m_\phi,\, m_{\chi_{1,2}}, \, m_{\psi_{1,2}} \ll T$. {}For the parameters considered in the text, these conditions are well satisfied. We can then write the cross-section of each one of the processes as
\begin{equation} \label{thermal_cross_section}
\sigma_i\sim{|\mathcal{M}_i|^2\over 16\pi T^2}~,
\end{equation}
where $|\mathcal{M}_i|^2$ is the averaged amplitude square of the process. The rate is then given by $\Gamma_i=\langle\sigma_i n_i v\rangle$, where for relativistic particles we have $n_i=f_i\zeta(3) T^3/\pi^2$ (where $\zeta(3) \simeq 1.20206$), with $f_i$ denoting the number of degrees of freedom (e.g. $f_{\chi_i} = 2$ for the complex scalars $\chi_{1,2}$, etc), and $v\sim 1$.
Hence, at leading-order, we find the tree-level estimates valid in the relativistic regime $T\gg m_{\phi,\chi,\psi}$:
\begin{align}
\Gamma_{\psi_i\psi_i  \leftrightarrows \delta\phi\delta\phi} &\simeq \sum_{i=1,2} \frac{ \zeta(3)}{4\pi^3}\left(y_i^4 T+ 2\lambda_{\psi_i}^2T^3 \right) \simeq   \frac{ \zeta(3)}{16\pi^3}   g_{\phi\psi}^2 T\left( 3 g_{\phi\psi}^2 + 2 \frac{T^2}{M^2} \right) , 
\label{rate1}\\
\Gamma_{\delta\phi\psi_i \leftrightarrows\delta\phi\psi_i} &\simeq \sum_{i=1,2}  \frac{3\zeta(3)}{8\pi^3} \left( y_i^4T  + \lambda_{\psi_i}^2T^3  \right)  \simeq   \frac{ \zeta(3)}{16\pi^3}   g_{\phi\psi}^2 T\left( 3 g_{\phi\psi}^2 + \frac{T^2}{M^2} \right), 
\label{rate2} \\
\Gamma_{\chi_i\chi_i \leftrightarrows \delta\phi\delta\phi}  &\simeq \sum_{i=1,2} \frac{\zeta(3)}{4\pi^3}T\left(\Lambda_{\chi_i}^2 + \frac{\kappa_i^4}{T^4}\right) \simeq \frac{\zeta(3)}{4\pi^3} g_{\phi\chi}^4 T \left(  1+ g_{\phi\chi}^4 \frac{M^4}{T^4} \right), 
\label{rate3} \\
\Gamma_{\delta\phi\chi_i  \leftrightarrows \delta\phi\chi_i} &\simeq \sum_{i=1,2} \frac{\zeta(3)}{4\pi^3}T\left(\Lambda_{\chi_i}^2 + \frac{\kappa_i^4}{T^4}\right)\simeq \frac{\zeta(3)}{4\pi^3} g_{\phi\chi}^4 T \left(  1+ g_{\phi\chi}^4 \frac{M^4}{T^4} \right),
\label{rate4}
\end{align}
where in the last step in the above equations, we have taken the average of the oscillatory terms using that $\phi \gg M$, which is valid for the model under consideration. {}For the parameters considered in section~\ref{results-WLI}, our only free parameter is the inflaton-$\chi$ coupling $g_{\phi\chi}$, which can then be used to control the magnitude of the total rate $\Gamma_{\rm scatt}$, given
by the sum of the rates (\ref{rate1})-(\ref{rate4}).

\begin{center}
\begin{figure}[!bth]
\centerline{\includegraphics[width=7.5cm]{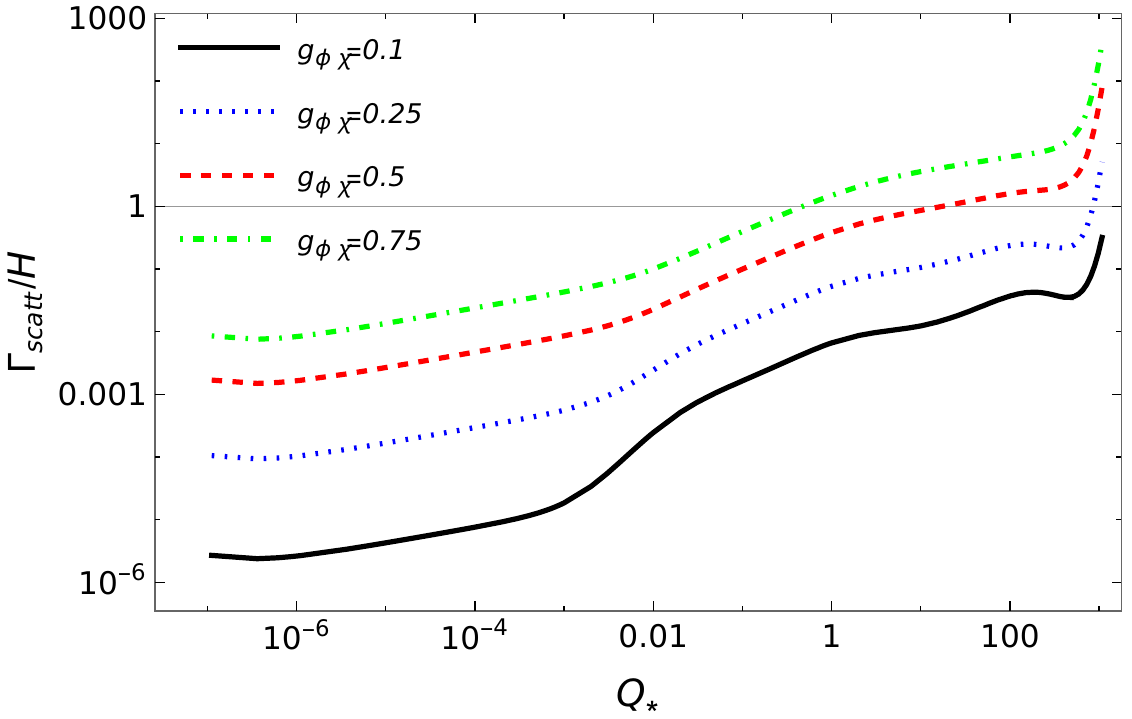}}
\caption{The total rate for the tree-level $2 \to 2$ processes 
divided by the Hubble rate in the case of $b=20$. The results are computed at the Hubble-exit point.}
\label{fig7}
\end{figure}
\end{center}

In fig.~\ref{fig7}, we compare the magnitude of the total rate $\Gamma_{\rm scatt}$ with the Hubble rate.  The results shown in fig.~\ref{fig7} are indicative that thermalization of the inflaton excitations can occur when the coupling $g_{\phi\chi}$ is sufficiently large and/or in the strong dissipative regime $Q \gg 1$. These results should also be seen as a lower-order estimate since we are only including lowest order tree-level processes and not including loop-mediated processes, which can be of same order, and also inelastic
processes like $2\to 3$,  $2\to 4$, etc. All of these processes will contribute to the thermalization of the inflaton excitations. 


\end{document}